\documentclass[preprint,superscriptaddress,pra]{revtex4}%
\usepackage{amsmath,amssymb}
\usepackage{graphicx}

\providecommand{\U}[1]{\protect\rule{.1in}{.1in}}
\ifx\pdfoutput\relax\let\pdfoutput=\undefined\fi
\newcount\msipdfoutput
\ifx\pdfoutput\undefined\else
\ifcase\pdfoutput\else
\ifx\paperwidth\undefined\else
\ifdim\paperheight=0pt\relax\else\pdfpageheight\paperheight\fi
\ifdim\paperwidth=0pt\relax\else\pdfpagewidth\paperwidth\fi
\fi\fi\fi

\begin{document}
\title{Control of quantum dynamics by optimized measurements}
\author{Feng Shuang}
\affiliation{Department of Chemistry, Princeton University, Princeton, New Jersey 08544}
\author{Mianlai Zhou}
\affiliation{National Laboratory for Physical Sciences at Micro Scale, University of
Science and Technology of China, Hefei, China}
\author{Alexander Pechen}
\affiliation{Department of Chemistry, Princeton University, Princeton, New Jersey 08544}
\author{Rebing Wu}
\affiliation{Department of Chemistry, Princeton University, Princeton, New Jersey 08544}
\author{Ofer M. Shir}
\affiliation{Department of Chemistry, Princeton University, Princeton, New Jersey 08544}
\affiliation{Natural Computing Group, Leiden University, Leiden, Netherlands}
\author{Herschel Rabitz}
\affiliation{Department of Chemistry, Princeton University, Princeton, New Jersey 08544}

\begin{abstract}
Quantum measurements are considered for optimal control of quantum dynamics
with instantaneous and continuous observations utilized\ to manipulate
population transfer. With an optimal set of measurements, the highest yield in
a two-level system can be obtained. The analytical solution is given for the
problem of population transfer by measurement-assisted coherent control in a
three-level system with a dynamical symmetry. The anti-Zeno effect is
recovered in the controlled processes. The demonstrations in the paper show
that suitable observations can be powerful tools in the manipulation of
quantum dynamics.
\end{abstract}
\date{\today}
\maketitle

\section{Introduction}

Control of quantum processes has attracted considerable attention, both
theoretically~\cite{Rice00,Rabitz00824,Rabitz0364,Shapiro03,
Bonacic-Koutecky05CR105, Alessandro2007} and
experimentally~\cite{Walmsley03PT56,Brixner011, Dantus04CR104}. Most studies of
quantum control are concerned with shaping a laser pulse to achieve an
expected goal. However, a coherent laser pulse is not the only driving force
for quantum dynamics. Incoherent driving forces, such as laser noise,
decoherence from the environment, and quantum observations, can also influence
quantum dynamics. A natural general expectation is that the later influence
will be deleterious toward achieving control~\cite{Zhu03JCP118}. However,
recent studies~\cite{Shuang04JCP121,Shuang06JCP124} have shown that controlled
quantum dynamics can survive intense field noise and decoherence, as well as
even cooperate with them under special circumstances~\cite{Shuang07PRE75}.
Under these special conditions, it is possible to meet the target goal more
effectively with the help of laser noise and environmental decoherence.
Incoherent non-unitary control of quantum systems by a suitably optimized
environment (e.g., incoherent radiation, a gas or solvent, a cloud of
electrons, atoms or molecules, etc.) was suggested as a supplement to coherent
control to provide a general tool for selective manipulation of both the
Hamiltonian and dissipative aspects of the system dynamics~\cite{Pechen06PRA73,
pechen2008}.

Both the outcome and back-action from quantum measurements could be used to
control quantum processes. In standard closed-loop optimal
control~\cite{Judson92PRL68}, the quantum system is non-selectively measured
when the desired evolution ends, and the outcomes of the measurements are used
by a learning algorithm to optimize the laser pulse. Measurements were also
used to map an unknown mixed state onto a known target pure
state~\cite{Roa06PRA73}. Some investigations considered exploiting the back
action from the quantum observations, and control by measurement plus
dynamical evolution was proposed~\cite{Vilela03PRA67}. The control of the
population branching ratio between two degenerate states by continuous
measurements was treated~\cite{Rice049984}, and the effect of nonoptimized
measurements on control by lasers was investigated~\cite{Sugawara05204115,
Sugawara06457}.

Numerical simulations have been performed to investigate observations serving
as indirect controls in the manipulation of quantum
dynamics~\cite{Shuang07JCP126}. Optimal control fields were shown to be capable
of cooperating or fighting with observations to achieve a good yield, and the
nature of the observations may be optimized to more effectively control the
quantum dynamics. Quantum observations also can break dynamical symmetries to
increase the controllability of a quantum system. The quantum Zeno and
anti-Zeno effects induced by observations are the key operating principles in
these processes. When instantaneous observations are the only forces to drive
population transfer in a two-level system, the maximal population transfer
induced by a finite number of measurements has been found, and the quantum
anti-Zeno effect is recovered in the limit of infinitely frequent
measurements~\cite{Pechen06PRA74}.

In this paper, we further explore the utility of quantum measurements as
controls in the manipulation of quantum dynamics. Here we assume that any
projection operator may be observed in a suitably performed experiment.
Analytical solutions and upper bounds of several controlled processes are
found. The remainder of the paper is organized as follows: Section II reviews
the main concepts of instantaneous and continuous measurements, which are used
as controls in this paper. The analytical solutions for instantaneous and
continuous observations in a two-level system are explored in Sec. III and IV,
respectively. The maximal measurement-assisted population transfer in a system
with dynamical symmetry is obtained in section V. A brief summary is presented
in section VI.

\section{Quantum Observations}

Quantum measurement serves as an incoherent driving force, and there are two
general types of quantum measurements: instantaneous von Neumann measurements
and continuous measurements. A characteristic feature of quantum systems is
that their measurement unavoidably affects the associated dynamics. The well
known manifestation of this back reaction is the uncertainty
principle~\cite{Mensky1993}. The influence of a measurement is revealed in a
more direct way through a change of the measured system state. In von Neumann
axiomatic quantum mechanics it is postulated that any measurement gives rise
to an abrupt change of the state of the system (instantaneous measurements)
under consideration and projects it onto an eigenstate of the measured
observable~\cite{VonNeumann1955}. The measurement process follows irreversible
dynamics, and causes the disappearance of coherence of the system's state: the
off-diagonal elements of the density matrix decay, or the phases of the
wavefunction amplitudes are randomized. Density matrices are adopted to
describe the states of controlled systems, because the nonselective quantum
measurements in the paper are applied to an ensemble of quantum systems.

\subsection{Instantaneous measurements}

An ideal measurement occurs at one instant of time or a sequence of such
observations may be performed at different instants~\cite{VonNeumann1955}. Let
$Q=\sum_{i}q_{i}P_{i}$ be an observable with $q_{i}$ being the eigenvalue of
$P_{i}$, which is a projector such that $P_{i}P_{j}=P_{i}\delta_{ij}$ and
$\sum_{i}P_{i}=1$. The measurement of $Q$ converts the state $\rho$ of the
system just before the measurement into the state
\begin{equation}
\rho^{\prime}\equiv\mu_{Q}\left(  \rho\right)  =\sum_{k}P_{k}\rho
P_{k}\text{.}\label{GM}
\end{equation}

A projection operator $P$ satisfies $P=P^{2}$, and its spectral decomposition
may be written as $P=q_{0}P_{0}+q_{1}P_{1}$ with the two eigenvalues being
$q_{0}=0$ and $q_{1}=1$ and two corresponding projectors being $P_{0}=1-P$ and
$P_{1}=P$. Thus, according to Eq.~(\ref{GM}), observation of the operator
$Q=P$ transforms the density matrix $\rho$ of the system into $\rho^{\prime}$
given by
\begin{subequations} \label{InstM}
\begin{align}
\rho^{\prime}  &  =\mu_{P}\left(  \rho\right)  =P_{0}\rho P_{0}+P_{1}\rho
P_{1}\\
&  =\rho-\left[  P,\left[  P,\rho\right]  \right]  \text{,}
\end{align}
\end{subequations}
so $\left[  P,\left[  P,\rho\right]  \right]  $ is the "kick" by an
instantaneous observation of the projection operator $P$.

\subsection{Continuous measurements}

There are two equivalent theoretical formulations of continuous quantum
measurements~\cite{Mensky94159}. One of them is based on restricted path
integrals (RPI) and the other one on master equations (ME). For simplicity, we
adopt the latter formulation. Corresponding to a continuous measurement of a
single observable, the master equation has the form~\cite{Walls1994}:
\begin{equation}
\dot{\rho}=-i\left[  H,\rho\right]  -\frac{1}{2}\kappa\left[  A,\left[
A,\rho\right]  \right]  \text{.} \label{ME}%
\end{equation}
Here, $H$ is the Hamiltonian of the measured system, $A$ is the measured
operator. Equation~(\ref{ME}) is similar with the equation describing a system
interacting with the environment, so we could call $\kappa$ as the "strength"
of the observation.

\subsection{Quantum ZENO and Anti-ZENO Effect}

Prevention of a quantum system's time evolution by means of repetitive,
frequent observations or continuous observations of the system's state is the
quantum Zeno effect (QZE). The QZE was proposed by Misra and
Sudarshan~\cite{Misra77756} and was experimentally
demonstrated~\cite{Itano902295} in a repeatedly measured two-level system
undergoing Rabi oscillations. A time-dependent observable projection operator
inducing up to $100\%$ transfer from one state to another
state~\cite{Roy004019} is the quantum anti-Zeno effect (QAZE). The impacts of
QZE and QAZE operations are the key processes explored in this paper to help
control quantum dynamics.

\section{Two-level system controlled by instantaneous measurements}

This section presents the analytical solution for the control of population
transfer in a two-level system by optimized instantaneous measurements. The
objective is to drive the population from the initial state $\rho
_{0}=\left\vert 0\right\rangle \left\langle 0\right\vert $ to the final state
$\rho_{f}=\left\vert 1\right\rangle \left\langle 1\right\vert $. An observable
$Q$ has the form $Q=q_{1}P_{1}^{\left(  Q\right)  }+q_{2}P_{2}^{\left(
Q\right)  }$, where $q_{1}$ and $q_{2}$ its the eigenvalues, and
$P_{1}^{\left(  Q\right)  }$ and $P_{2}^{\left(  Q\right)  }$ are the
corresponding projectors. From Eq.~(\ref{InstM}), it's easy to see that the
measurement of $Q$ is equal to the measurement of the projector $P_{1}%
^{\left(  Q\right)  }$, since $P_{2}^{\left(  Q\right)  }=1-P_{1}^{\left(
Q\right)  }$. Thus, it is sufficient to consider the measurement of projection
operators in this case.

A sequence of $N$ instantaneous projection observations, specified by the
operators
\begin{subequations}
\begin{align}
P_{k}  &  =\left\vert \psi_{k}\right\rangle \left\langle \psi_{k}\right\vert
\text{, }\label{Pk}\\
\left\vert \psi_{k}\right\rangle  &  =\cos\frac{\alpha_{k}}{2}\left\vert
0\right\rangle +e^{i\theta_{k}}\sin\frac{\alpha_{k}}{2}\left\vert
1\right\rangle \text{.} \label{Psik}%
\end{align}
\end{subequations}
are performed at times $T_{k}$, $k=1,\cdots,N$. Parameters $\alpha_{k}$ and
$\theta_{k}$ in Eq.~(\ref{Psik}) are limited to the range%
\begin{subequations}
\begin{align}
-\frac{\pi}{2}  &  \leq\frac{\alpha_{k}}{2}<\frac{\pi}{2}\text{,
}\label{ARange}\\
0  &  \leq\theta_{k}<\pi\text{,} \label{TRange}%
\end{align}
\end{subequations}
since the projection operator $P_{k}$ does not depend on the phase of
$\left\vert \psi_{k}\right\rangle $. The operators $P_{k}$ parameterized by
$\alpha_{k}$ and $\theta_{k}$, $k=1,\cdots,N$, are optimized to maximize the
yield,%
\begin{equation}
Y_{N}\left[  P_{1,}\cdots,P_{N}\right]  =\left\langle 1\right\vert \rho
_{N}\left\vert 1\right\rangle \text{.} \label{YNDef}%
\end{equation}
Here $\rho_{N}$ is the density matrix after performance of $N$ observations
given by the iterative equation
\begin{equation}
\rho_{k}=\rho_{k-1}-\left[  P_{k},\left[  P_{k},\rho_{k-1}\right]  \right]
\text{,}\qquad k=1,\cdots,N \label{RhoIte}%
\end{equation}
with $\rho_{0}=\left\vert 0\right\rangle \left\langle 0\right\vert $ and
$P_{k}$ described in Eq.~(\ref{Pk}). We have neglected the free evolution
between measurements, which is easy to include via a transformation between
the Schr\"{o}dinger picture and interaction picture,
\begin{subequations}
\begin{align}
\rho_{k}  &  =e^{iH_{0}T_{k}}\rho_{k}^{\left(  S\right)  }e^{-iH_{0}T_{k}%
}\text{,}\\
P_{k}  &  =e^{iH_{0}T_{k}}P_{k}^{\left(  S\right)  }e^{-iH_{0}T_{k}}\text{.}%
\end{align}
\end{subequations}
Here $\rho_{k}^{\left(  S\right)  }$ is the density matrix in the
Schr\"{o}dinger picture governed by the iterative equation,
\begin{equation}
\rho_{k}^{\left(  S\right)  }=e^{-iH_{0}\left(  T_{k}-T_{k-1}\right)  }%
\{\rho_{k-1}^{\left(  S\right)  }-[P_{k}^{\left(  S\right)  },[P_{k}^{\left(
S\right)  },\rho_{k-1}^{\left(  S\right)  }]]\}e^{iH_{0}\left(  T_{k}%
-T_{k-1}\right)  },k=1,\cdots,N\text{,}%
\end{equation}
which includes the free evolution.

The density matrix $\rho_{N}$ is Hermitian with unit trace. Hence, it can be
expressed in the for
\begin{equation}
\rho_{N}=\left(
\begin{array}
[c]{cc}%
1-Y_{N} & Z_{N}^{\ast}\\
Z_{N} & Y_{N}%
\end{array}
\right)  \text{.}%
\end{equation}
It is easy to establish the following solution to the iteration in Eq.
(\ref{RhoIte})
\begin{subequations}\label{XZ}
\begin{align}
Y_{k}  &  =\left\langle 1\right\vert \rho_{k}\left\vert 1\right\rangle
=\frac{1}{2}\left(  1-\cos\alpha_{1}C_{12}C_{23}\cdots C_{k-1,k}\cos\alpha
_{k}\right)  \text{,}\label{YN}\\
Z_{k}  &  =\left\langle 1\right\vert \rho_{k}\left\vert 0\right\rangle
=\frac{1}{2}e^{i\theta_{k}}\cos\alpha_{1}C_{12}C_{23}\cdots C_{k-1,k}%
\sin\alpha_{k}\text{,}
\end{align}
\end{subequations}
with the coefficients $C_{mn}$ given by
\begin{equation}
C_{mn}=\cos\alpha_{m}\cos\alpha_{n}+\cos\left(  \theta_{m}-\theta_{n}\right)
\sin\alpha_{m}\sin\alpha_{n}\text{.}
\end{equation}
Therefore,
\begin{equation}
Y_{N}=\frac{1}{2}\left(  1-\cos\alpha_{1}C_{12}C_{23}\cdots C_{N-1,N}%
\cos\alpha_{N}\right)
\end{equation}
is the yield from $N$ observations.

We will now determine the maximum value of $Y_{N}$, which is a function of
variables $\theta_{k}$ and $\alpha_{k}$. The inequality~(\ref{TRange}) yields
$-\pi<\theta_{m}-\theta_{n}<\pi$ and $1\geq\cos\left(  \theta_{m}-\theta
_{n}\right)  >-1$. Setting to zero the derivative of $Y_{N}$ with respect to
$\theta_{k}$ gives $\sin\left(  \theta_{k}-\theta_{k-1}\right)  =0$ for
$k=2,\cdots,N$. Hence $Y_{N}$ reaches its maximum when%
\begin{equation}
\theta_{1}=\theta_{2}=\cdots=\theta_{N}%
\end{equation}
Therefore, after optimization with respect to $\theta_{k}$, $Y_{N}$ can be
written as an function of $\alpha_{k}$, $k=1,\cdots,N$,
\begin{subequations}
\begin{align}
Y_{N}^{\left(  \alpha\right)  }  &  =\frac{1}{2}\left[  1-\cos\alpha_{1}
\cos\left(  \alpha_{1}-\alpha_{2}\right)  \cdots\cos\left(  \alpha
_{N-1}-\alpha_{N}\right)  \cos\alpha_{N}\right] \label{YNAlpha}\\
&  =\frac{1}{2}\left[  1+\prod\nolimits_{k=0}^{N}\cos\varphi_{k}\right]
\text{,}%
\end{align}
\end{subequations}
with $\varphi_{0}=\pi-\alpha_{1}$, $\varphi_{1}=\alpha_{1}-\alpha_{2}$,
$\cdots$, $\varphi_{N-1}=\alpha_{N-1}-\alpha_{N}$ and $\varphi_{N}=\alpha_{N}%
$. It is easy to verify that the second derivative of the function $f\left(
x\right)  =\ln\cos\left(  x\right)  $ is negative, so it is a concave
function. The inequality,
\begin{equation}
\prod\nolimits_{k=0}^{N}\cos\varphi_{k}\leq\left(  \cos\frac{\pi}{N+1}\right)
^{N}\text{,}%
\end{equation}
can be established by the majorization inequality~\cite{Roberts1973} for
concave functions,%
\begin{equation}
\frac{\sum_{k=1}^{M}f\left(  x_{k}\right)  }{M}\leq f\left(  \frac{\sum
_{k=1}^{M}x_{k}}{M}\right)  \text{.}%
\end{equation}
Hence, $Y_{N}^{\left(  \alpha\right)  }$ reaches its maximum value%
\begin{equation}
Y_{N}^{\left(  O\right)  }=\frac{1}{2}\left[  1+\left(  \cos\frac{\pi}%
{N+1}\right)  ^{N+1}\right]  \text{,} \label{YIO}%
\end{equation}
when $\varphi_{0}=\cdots=\varphi_{N}=\frac{\pi}{N+1}$. The solutions are
consistent with what was found in~\cite{Pechen06PRA74}. The QAZE is recovered
in the limit of an infinite number of observations,%
\begin{equation}
\lim_{N\rightarrow\infty}Y_{N}^{\left(  O\right)  }=1\text{.}%
\end{equation}

\section{Two-level system controlled by continuous measurements}

In this section, the quantum dynamics of a two-level system is controlled by
suitable continuous measurements. Here we assume that it is possible to
continuously measure any time-dependent projection operator $P\left(
t\right)  $. In the interaction picture, the dynamics of the continuous
observation process is described by%
\begin{equation}
\dot{\rho}\left(  t\right)  =-\gamma\mathcal{L}\left(  t\right)  \rho\left(
t\right)  =-\gamma\left[  P\left(  t\right)  ,\left[  P\left(  t\right)
,\rho\left(  t\right)  \right]  \right]  \text{,} \label{Lrho}%
\end{equation}
where $\mathcal{L}\left(  t\right)  $ is a super-operator acting on the
density matrix $\rho\left(  t\right)  $, and $\gamma$ is the constant strength
of the observation. The projection operator $P\left(  t\right)  $ is specified
by
\begin{subequations}
\label{CMOperator}%
\begin{align}
P\left(  t\right)   &  =\left\vert \psi\left(  t\right)  \right\rangle
\left\langle \psi\left(  t\right)  \right\vert \text{,}\label{CMP}\\
\left\vert \psi\left(  t\right)  \right\rangle  &  =\cos\frac{\alpha\left(
t\right)  }{2}\left\vert 0\right\rangle +e^{i\theta\left(  t\right)  }%
\sin\frac{\alpha\left(  t\right)  }{2}\left\vert 1\right\rangle \text{,}
\label{CMPsi}%
\end{align}
\end{subequations}
where $\alpha\left(  t\right)  $, $\theta\left(  t\right)  $ are functions of
time $t$ to be determined. The goal is to optimize the objective functional,
as the yield at final time $T_{f}$,%
\begin{equation}
Y\left(  T_{f}\right)  \left[  P\left(  t\right)  \right]  =Y\left(
T_{f}\right)  \left[  \alpha\left(  t\right)  ,\theta\left(  t\right)
\right]  =\left\langle 1\right\vert \rho\left(  T_{f}\right)  \left\vert
1\right\rangle \text{,} \label{JP}%
\end{equation}
where the system is initially populated on state $\left\vert 0\right\rangle $.

Eq. (\ref{Lrho}) appears insoluble for general functions of $\alpha\left(
t\right)  $ and $\theta\left(  t\right)  $. First consider only a simple case,
with zero phase and $\alpha\left(  t\right)  $ taken as linear in time,
\begin{subequations}
\label{Assumption}%
\begin{align}
\alpha\left(  t\right)   &  =A\frac{t}{T_{f}}+B\\
\theta\left(  t\right)   &  =0\text{.}%
\end{align}
\end{subequations}
The final yield of this case may be explicitly worked out as%
\begin{equation}
Y\left(  T_{f}\right)  =\frac{1}{2}-\frac{1}{2}e^{-\gamma^{\prime}}\left\{
\cos A\cosh\delta+\left[  \gamma^{\prime}\cos\left(  2B+A\right)  +A\sin
A\right]  \frac{\sinh\delta}{\delta}\right\}  \text{,} \label{yieldsol}%
\end{equation}
where $\gamma^{\prime}$ and $\delta$ are dimensionless parameters defined by
\begin{subequations}
\begin{align}
\gamma^{\prime}  &  =\frac{1}{2}\gamma T_{f}\label{gammap}\\
\delta &  =\sqrt{\gamma^{\prime2}-A^{2}}\text{.}%
\end{align}
\end{subequations}
Eq. (\ref{yieldsol}) reaches its maximum value when
\begin{subequations}
\begin{align}
2B_{m}+A_{m}  &  =\pi\\
\gamma^{\prime}\sin A_{m}  &  =A_{m}\text{,} \label{gammaA}%
\end{align}
\end{subequations}
in which case the optimal value of the yield is
\begin{equation}
Y^{\left(  O\right)  }\left(  T_{f}\right)  =\frac{1}{2}\left(  1-e^{-\gamma
^{\prime}\left(  1+\cos A_{m}\right)  }\cos A_{m}\right)  \text{.} \label{YCO}%
\end{equation}
Figure 1 depicts the variation of the optimal $A_{m}$ and $B_{m}$ with respect
to $\gamma^{\prime}$, and we conclude that
\begin{subequations}
\begin{align}
A_{m}  &  =0\text{, for }\gamma^{\prime}<1\\
A_{m}  &  \rightarrow\pi\text{, for }\gamma^{\prime}\rightarrow\infty\text{.}%
\end{align}

It follows from Eq.~(\ref{YCO}) that a complete population transfer is
attained when $\gamma^{\prime}$ in Eq.~(\ref{gammap}) approaches infinity.
Thus, from Eq.~(\ref{gammap}), increasing the observation strength $\gamma$
and the final time are equally effective in enhancing the control process, and
the QAZE is recovered in the limit of infinite observation strength, or final time.

We now assess whether the linear solution in Eq.~(\ref{Assumption}) is optimal
with respect to all possible forms of $\alpha\left(  t\right)  $ and
$\theta\left(  t\right)  $. To verify that this is the case, we start from Eq.~(\ref{Lrho}) and consider the variation of $\rho\left(  t\right)  $ with
respect to $\alpha\left(  t\right)  $ and $\theta\left(  t\right)  $. The
general variation of Eq.~(\ref{Lrho}) gives
\end{subequations}
\begin{equation}
\frac{d\left[  \delta\rho\left(  t\right)  \right]  }{dt}=-\gamma\left[
\mathcal{L}\left(  t\right)  \delta\rho\left(  t\right)  +\delta
\mathcal{L}\left(  t\right)  \rho\left(  t\right)  \right]  \text{.}%
\end{equation}
and for a driving variation $\delta\alpha\left(  t\right)  $, it is easy to
verify that the solution of the above equation is%
\begin{subequations}
\begin{align}
\delta\rho\left(  t\right)   &  =-\gamma\int_{0}^{t}\left[  \rho_{\alpha
}\left(  t,\tau\right)  \right]  \delta\alpha\left(  \tau\right)
d\tau\text{,}\\
\rho_{\alpha}\left(  t,\tau\right)   &  =\mathcal{U}\left(  t,\tau\right)
[\frac{d\mathcal{L}\left(  \tau\right)  }{d\alpha}\rho\left(  \tau\right)
]\text{,}%
\end{align}
where $\mathcal{U}\left(  T,t\right)  $ is a time-ordered exponential%
\end{subequations}
\begin{equation}
\mathcal{U}\left(  t,\tau\right)  =\exp_{+}\left[  -\gamma\int_{\tau}%
^{t}\mathcal{L}\left(  \nu\right)  d\nu\right]  \text{.}%
\end{equation}
Hence $\rho_{\alpha}\left(  t,\tau\right)  $ is the solution of the
differential equation%
\begin{equation}
\frac{\partial\rho_{\alpha}\left(  t,\tau\right)  }{\partial t}=-\gamma
\mathcal{L}\left(  t\right)  \rho_{\alpha}\left(  t,\tau\right)
\end{equation}
with the initial condition
\begin{equation}
\rho_{\alpha}\left(  \tau,\tau\right)  =\frac{d\mathcal{L}\left(  \tau\right)
}{d\alpha}\rho\left(  \tau\right)  \text{.}%
\end{equation}
It is evident that $\rho_{\alpha}\left(  t,\tau\right)  $ is real, symmetric
and traceless under the assumption of Eq.~(\ref{Assumption}), hence we can
set
\begin{equation}
\rho_{\alpha}\left(  t,\tau\right)  =\left(
\begin{array}
[c]{cc}%
-Y_{\alpha}\left(  t,\tau\right)  & Z_{\alpha}\left(  t,\tau\right) \\
Z_{\alpha}\left(  t,\tau\right)  & Y_{\alpha}\left(  t,\tau\right)
\end{array}
\right)  \text{.}%
\end{equation}
It is easy to yield the result%
\begin{equation}
Y_{\alpha}\left(  t,\tau\right)  =-\frac{1}{2}\sin A\left(  1-\frac{t}{T_{f}%
}\right)  \exp\left[  \frac{A\left(  t-2\tau\right)  \cot A-t\csc A}{T_{f}%
}\right]  \text{.} \label{ccfollow2}%
\end{equation}
So the variation of the final yield with respect to $\alpha\left(  t\right)  $
is zero,
\begin{equation}
\delta Y^{\left(  O\right)  }\left(  T_{f}\right)  =-\gamma\int_{0}^{T_{f}%
}Y_{\alpha}\left(  T_{f},\tau\right)  \delta\alpha\left(  \tau\right)
dt=0\text{.}%
\end{equation}
Using the same procedures taken above, we can prove that the variation of the
final yield with respect to the phase function $\theta\left(  t\right)  $ is
also zero. Hence, the linear solution is an optimal solution.

In addition to the analysis above, we performed numerical simulations, where
the goal was optimization of the yield $Y$ by means of an evolutionary
algorithm approach to determine $\alpha(t)$ and $\theta(t)$. The optimization
procedure was conducted freely, without any preliminary assumptions on
$\alpha(t)$ or $\theta(t)$, and without any constraints on their values during
the search. For this purpose, we applied the covariance matrix adaptation
evolution strategy (CMA-ES)~\cite{Hansen01completely,hansencmamultimodal} to
the task. The latter algorithm is very efficient for treating continuous
global optimization problems~\cite{HansenCEC2005b, SHIR_CEC06}. It has been
successful for handling correlations among object variables. Fig. 2 depicts
the best yield from performing full optimization with respect to
$\alpha\left(  t\right)  $ and $\theta\left(  t\right)  $, when the
assumptions in Eq.~(\ref{Assumption}) are not applied. It can be concluded
that the solution is globally optimal.

The best yield from continuous measurements, Eq.~(\ref{YCO}), seems very
different to the best yield from instantaneous measurements, Eq.~(\ref{YIO}).
However, their asymptotic forms,
\begin{subequations}
\begin{align}
Y_{N}^{\left(  O\right)  }  &  \thicksim1-\frac{\pi^{2}}{4N}\,,\text{
\ \ }N\rightarrow\infty\text{,}\\
Y^{\left(  O\right)  }\left(  T_{f}\right)   &  \thicksim1-\frac{\pi^{2}%
}{2\gamma T_{f}},\text{ }\gamma T_{f}\rightarrow\infty\text{,}%
\end{align}
\end{subequations}
are very similar. Hence the best yield from continuous measurements with
measurement strength $\gamma$ is very close to the best yield from a sequence
of $N\simeq\gamma T_{f}/2$ instantaneous measurements, when $\gamma T_{f}\gg1$.

\section{Optimal population transfer in a system with dynamical symmetry by
measurement-assisted coherent control}

In this section we consider a system whose free Hamiltonian $H_{0}$ and dipole
moment $\mu$ are given by
\[
H_{0}=\left(
\begin{array}
[c]{ccc}%
1 & 0 & 0\\
0 & 2 & 0\\
0 & 0 & 3
\end{array}
\right)  ,\qquad\mu=\left(
\begin{array}
[c]{ccc}%
0 & 1 & 0\\
1 & 0 & 1\\
0 & 1 & 0
\end{array}
\right)  .
\]
The system is initially prepared in the ground state $|\psi_{0}\rangle
=|0\rangle$ at $t=0$. The control goal is to transfer as much as possible of
the population from the ground state $|0\rangle$ to the first excited state
$|1\rangle$ at the target time $T>0$ using as controls a coherent
electromagnetic field $\varepsilon(t)$ during the time period $[0,T]$ and a
single measurement of the projector $P_{0}=|0\rangle\langle0|$ (or
$P_{2}=|2\rangle\langle2|$) at a time $t_{1}\in(0,T)$.

The symmetry of the system implies (see Ref.~\cite{Rabitz011}) that the
coefficients of a pure system state $|\psi_{t}\rangle=C_{0}(t)|0\rangle
+C_{1}(t)|1\rangle+C_{2}(t)|2\rangle$ satisfy the following relation upon
evolution under only the action of a coherent field,
\begin{equation}
\left\vert C_{0}(t)C_{2}(t)-\frac{C_{1}^{2}(t)}{2}\right\vert =\left\vert
C_{0}(0)C_{2}(0)-\frac{C_{1}^{2}(0)}{2}\right\vert \text{,}\quad \text{for any }t\geq0\text{.}
\label{eq0}%
\end{equation}
If the initial state is $|\psi_{0}\rangle=|0\rangle$, then $C_{1}%
(0)=C_{2}(0)=0$ and Eq.~(\ref{eq0}) becomes the following relation for the
coefficients $C_{i}(t)$:
\begin{equation}
C_{1}^{2}(t)=2C_{0}(t)C_{2}(t) \text{ ,}\quad \text{for any }t\geq0 \text{.}\label{eq1}%
\end{equation}
This relation was used in~\cite{Shuang07JCP126} to conclude that transferring
more than $50\%$ of the population from the level $|0\rangle$ to the level
$|1\rangle$ is impossible using only a coherent control field.

Measurements performed on the system can break the dynamical symmetry thereby
allowing for exceeding the above $50\%$ population transfer limitation.
Numerically, the measurement-assisted control problem for this system was
investigated in Ref.~\cite{Shuang07JCP126}, where transferring $66.9\%$ of the
population to the level $|1\rangle$ was obtained with a coherent control field
assisted by a single measurement of $P_{0}$. In this section we analytically
treat this control problem to find the upper bound on the maximal population
transfer to the level $|1\rangle$, which is found to be approximately
$68.7\%$. We also explicitly find the Rabi Frequencies of the optimal pulses,
thus providing a complete analytical solution to this problem.

The control process consists of the following three steps. First, the system
evolves under the action of a coherent field during the time interval
$[0,t_{1})$. Second, at the time $t=t_{1}$ a non-selective measurement of
$P_{0}$ is performed on the system, which transforms the system state in
accordance with the von Neumann scheme. Third, the system evolves during the
time interval $(t_{1},T]$ again only under the action of a coherent field.

Spontaneous emission during the first and third steps is neglected in this
consideration. Therefore the system's dynamics under the action of an
electromagnetic coherent field during the first and third steps can be
described by optical Bloch's equation without relaxation terms:%
\begin{equation}
\frac{d\rho(t)}{dt}=-i[H,\rho(t)]\text{.} \label{eq7}%
\end{equation}
Here the Hamiltonian $H=\Omega(t)|0\rangle\langle1|+\Omega(t)|1\rangle
\langle2|+\mathrm{h.c.}$ is determined by the Rabi frequency $\Omega(t)$ of
the electromagnetic field $\varepsilon(t)$. The symmetry of the system implies
that the Rabi frequencies for the transitions $|0\rangle\leftrightarrow
|1\rangle$ and $|1\rangle\leftrightarrow|2\rangle$ are the same.

Without loss of generality, it is sufficient to consider, constant Rabi
frequencies during each step of the control. If the Rabi frequency $\Omega$ is
time independent, then the Hamiltonian has the form $H\equiv H(\Omega
)=\Omega|0\rangle\langle1|+\Omega|1\rangle\langle2|+\mathrm{h.c.}$ and the
solution of~(\ref{eq7}) with the initial condition $\rho(t_{0})=\rho_{0}$ is
$\rho(t)=U(t-t_{0})\rho_{0}U^{\dagger}(t-t_{0})$, where $U(\tau)=e^{-i\tau
H(\Omega)}$. The Hamiltonian $H(\Omega)$ can be written in terms of the vector
$|\Omega\rangle=[\Omega|0\rangle+\Omega^{\ast}|2\rangle]/\sqrt{2}|\Omega|$ as
$H(\Omega)=\sqrt{2}|\Omega|[|1\rangle\langle\Omega|+|\Omega\rangle\langle1|]$
(here $\Omega^{\ast}$ is the complex conjugate of $\Omega$). One has
\[
\left\{
\begin{array}
[c]{l}%
\left[  H(\Omega)\right]  ^{2}=2|\Omega|^{2}[|1\rangle\langle1|+|\Omega
\rangle\langle\Omega|]\\
\left[  H(\Omega)\right]  ^{3}=2|\Omega|^{2}H
\end{array}
\right.  \Rightarrow\left\{
\begin{array}
[c]{l}%
\left[  H(\Omega)\right]  ^{2n}=(\sqrt{2}|\Omega|)^{2n}[|1\rangle
\langle1|+|\Omega\rangle\langle\Omega|]\\
\left[  H(\Omega)\right]  ^{2n+1}=(\sqrt{2}|\Omega|)^{2n+1}[|1\rangle
\langle\Omega|+|\Omega\rangle\langle1|]
\end{array}
\right.
\]
This gives
\begin{equation}
U(\tau)=P_{\widetilde{\Omega}}+\cos\Bigl(\sqrt{2}|\Omega|\tau
\Bigr)\Bigl(|1\rangle\langle1|+|\Omega\rangle\langle\Omega|\Bigr)-i\sin
\Bigl(\sqrt{2}|\Omega|\tau\Bigr)\Bigl(|1\rangle\langle\Omega|+|\Omega
\rangle\langle1|\Bigr) \label{eq8}%
\end{equation}
where $P_{\widetilde{\Omega}}=\mathbb{I}-|1\rangle\langle1|-|\Omega
\rangle\langle\Omega|=|\widetilde{\Omega}\rangle\langle\widetilde{\Omega}|$ is
the projector onto the subspace generated by the vector $|\widetilde{\Omega
}\rangle=(\Omega|0\rangle-\Omega^{\ast}|2\rangle)/\sqrt{2}|\Omega|$.

In the first stage of control, the initial density matrix $\rho_{0}$ is
transformed into $\rho_{1}=U_{1}\rho_{0}U_{1}^{\dagger}=|\psi_{1}%
\rangle\langle\psi_{1}|$, where $|\psi_{1}\rangle=U_{1}|0\rangle$ and
$U_{1}=\exp[-it_{1}H(\Omega_{1})]$ is the evolution operator induced by the
control field with some Rabi frequency $\Omega_{1}=|\Omega_{1}|e^{i\psi_{1}}$.
Direct calculations give $|\psi_{1}\rangle=C_{0}|0\rangle+C_{1}|1\rangle
+C_{2}|2\rangle$ with
\[
C_{0}=\frac{\cos\left(  \sqrt{2}|\Omega_{1}|t_{1}\right)  +1}{2},\qquad
C_{1}=\frac{i\sin\left(  \sqrt{2}|\Omega_{1}|t_{1}\right)  }{\sqrt{2}%
}e^{-i\psi_{1}},\qquad C_{2}=\frac{\cos\left(  \sqrt{2}|\Omega_{1}%
|t_{1}\right)  -1}{2}e^{-2\psi_{1}}%
\]

Measuring the projector $P_{0}$ at the time $t=t_{1}$ transforms the pure
state $\rho_{1}$ into the density matrix
\begin{align*}
\rho_{2}  &  =\mu_{P_{0}}(\rho_{1})=P_{0}\rho_{1}P_{0}+(\mathbb{I}-P_{0}%
)\rho_{1}(\mathbb{I}-P_{0})\\
&  =|C_{0}|^{2}P_{0}+|C_{1}|^{2}P_{1}+|C_{2}|^{2}P_{2}+C_{1}C_{2}^{\ast
}|1\rangle\langle2|+C_{1}^{\ast}C_{2}|2\rangle\langle1|
\end{align*}
If $C_{1}\neq0$, then the state $\rho_{2}$ is mixed and the measurement
destroys the coherence between the levels $|0\rangle$ and $|1\rangle$ while
preserving the coherence between $|1\rangle$ and $|2\rangle$.

After the measurement the system density matrix evolves under action of a
coherent field with some Rabi frequency $\Omega_{2}=|\Omega_{2}|e^{i\psi_{2}}$
into $\rho_{3}=U_{2}\rho_{2}U_{2}^{\dagger}$. Here $U_{2}=\exp[-it_{2}%
H(\Omega_{2})]$ is the evolution operator induced by the coherent field and
$t_{2}=T-t_{1}$. The density matrix $\rho_{3}=U_{2}\left[  \mu_{P_{0}}\left(
U_{1}\rho_{0}U_{1}^{+}\right)  \right]  U_{2}^{\dagger}$ can be computed
using~(\ref{eq8}). The computation gives the following population
$P=\langle1|\rho_{3}|1\rangle$ of level $|1\rangle$ at the target time $t=T$,
\begin{equation}
P=\frac{1}{16}\biggl\{5-\cos(x_{1})-[1+3\cos(x_{1})]\cos(x_{2})+2[2\sin
(x_{1}/2)-\sin(x_{1})]\sin(x_{2})\cos(\psi_{2}-\psi_{1})\biggr\}, \label{eq6}%
\end{equation}
where $x_{1}=2\sqrt{2}|\Omega_{1}|t_{1}$ and $x_{2}=2\sqrt{2}|\Omega_{2}%
|t_{2}$. This function is maximized by
\begin{subequations}
\begin{align}
x_{1}^{\ast}  &  =\pm\left[  2\arctan\left(  \frac{\sqrt{18+2\sqrt{6}}}%
{\sqrt{6}-1}\right)  -2\pi\right] \\
x_{2}^{\ast}  &  =\mp\arctan\left(  \frac{\sqrt{18+2\sqrt{6}}}{\sqrt{6}%
-1}\right)
\end{align}
and by $\psi_{1},\psi_{2}$ such that $\psi_{2}-\psi_{1}=2\pi k$, $k=0,\pm
1,\pm2,\dots$. The maximal value is
\end{subequations}
\begin{equation}
P_{\mathrm{max}}=\max\limits_{\Omega_{1},\Omega_{2}}P=4\cdot10^{-3}\left(
\sqrt{393-48\sqrt{6}}+138+7\sqrt{6}\right)  \approx68.7\%\text{.}
\label{P3max}%
\end{equation}
This maximal population transfer to the level $|1\rangle$ can be obtained by
applying a coherent field with Rabi frequency $\Omega_{1}=x_{1}^{\ast}%
/(2\sqrt{2}t_{1})e^{i\psi_{1}}$ to the system during the time interval
$\left[  0,t_{1}\right]  $, then measuring the projector $P_{0}$ at time
$t_{1}$, and finally applying a coherent field with Rabi frequency $\Omega
_{2}=x_{2}^{\ast}/(2\sqrt{2}t_{2})e^{i\psi_{2}}$ during the time interval
$\left[  t_{1},T\right]  $, where $\psi_{2}=\psi_{1}+2\pi k$, $k=0,\pm
1,\pm2,\dots$.

Another simple way to compute the maximal yield is with the well-known Euler
decomposition of the $SU(2)$ Lie group~\cite{Biedenharn1981}. Suppose the
system' propagation consists of the following three steps:
\begin{subequations}
\begin{align}
\rho_{1}  &  =U_{1}^{\dagger}\rho_{0}U_{1}\\
\rho_{2}  &  =\rho_{1}-\left[  P_{k},\left[  P_{k},\rho_{1}\right]  \right] \\
\rho_{3}  &  =U_{2}^{\dagger}\rho_{2}U_{2}%
\end{align}
where Euler's decomposition of the unitary propagators is as follows:
\end{subequations}
\begin{equation}
U_{k}=\exp\left(  ia_{k}H_{0}\right)  \exp\left(  i\frac{x_{k}}{2\sqrt{2}}%
\mu\right)  \exp\left(  ib_{k}H_{0}\right)  \text{, }k=1,2\text{,}%
\end{equation}
and $a_{1,2},b_{1,2}$ and $x_{1,2}$ are six independent variables to be
optimized with. Simple computation yields the population of the level
$|1\rangle$ as%
\begin{equation}
\left(  \rho_{3}\right)  _{11}=\frac{1}{16}\left[  5-\cos x_{2}-\cos
x_{1}\left(  1+3\cos x_{2}\right)  +2\cos\left(  a_{2}+b_{1}\right)  \left(
\sin x_{1}-2\sin\frac{x_{1}}{2}\right)  \sin x_{2}\right]  \text{,}%
\end{equation}
with which it is easy to derive the same maximal population transfer shown in
Eq.~(\ref{P3max}).

\section{Conclusion}
This paper discusses the use of both instantaneous and continuous
observations in the manipulation of quantum dynamics. The measurements can be
viewed as direct controls. Two-level systems and a special three-level system
are treated analytically. Solutions and upper bounds for the controlled
processes are obtained, and they agree very well with previous numerical
simulations, and QAZE is recovered. The results are proper for instantaneous
observations performed any number of times and continuous observations
performed with any strength. The performance of optimal observations hopefully
will become routine with advancing technology, as observations can be powerful
tools in the control of quantum dynamics.

\begin{acknowledgments}
The authors acknowledge support from the NSF and an ARO grant.
\end{acknowledgments}

%\bibliographystyle{apsrev}
%\bibliography{OptMeasure}

\pagebreak
\begin{figure}\center
\includegraphics[scale=1.5]{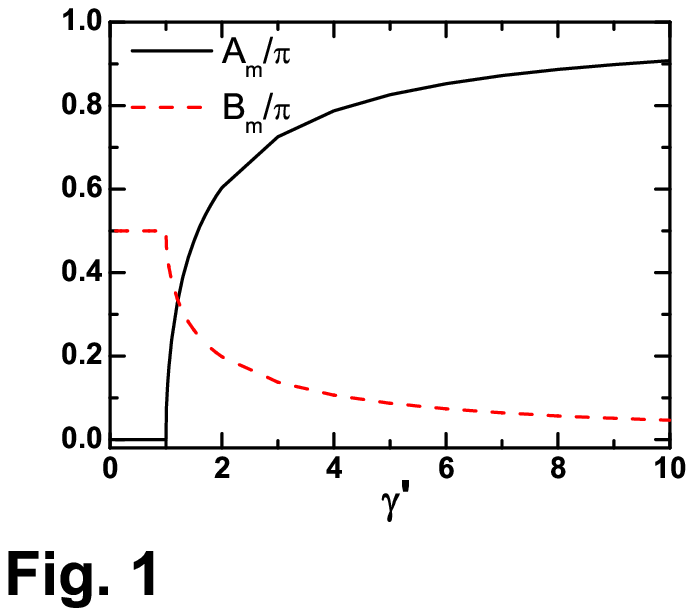}
\caption{(Color online) Optimal coefficients $A_{m}$ and $B_{m}$ of the
linear function $\alpha\left(  t\right)  $ in the projection operators (ref.
Eq.~(\ref{CMP})) which are continuously measured to control the quantum
dynamics of a two-level system. In the optimization process, the functions
$\alpha\left(  t\right)  $ and $\theta\left(  t\right)  $ in the projection
operators are assumed to be linear and zero, respectively, and the optimal
solutions are proved to be globally optimal. $\gamma^{\prime}$ (dimensionless)
is multiplication of observation strength and observation time (ref. Eq.~(\ref{gammap})).}
\end{figure}

\begin{figure}\center
\includegraphics[scale=1.5]{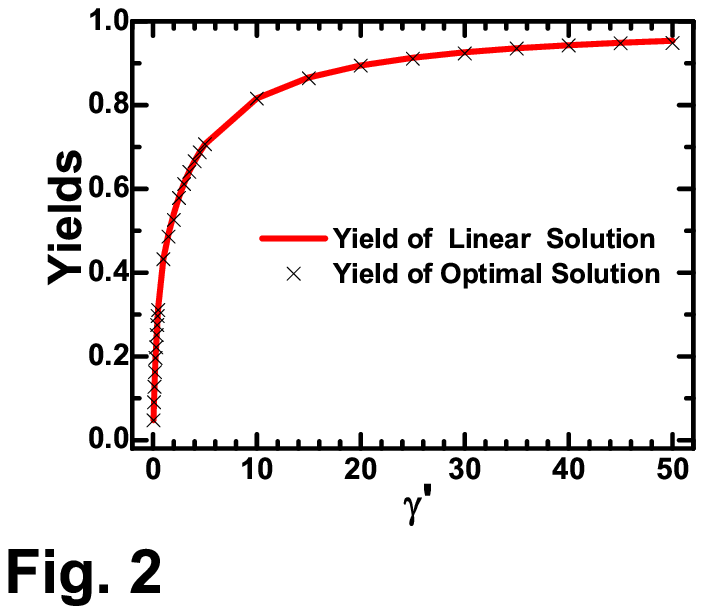}
\caption{(Color online) The yield as a function of the observation strength for
the linear solution (solid line), with the best attained yield with an
evolutionary search using the CMA-ES algorithm (squares) as a reference. The
latter non-linear solutions did not exceed the linear solution's global
optimal yield. $\gamma^{\prime}$ (dimensionless) is multiplication of
observation strength and observation time (ref. Eq.~(\ref{gammap})).}
\end{figure}

\end{document}